\documentclass[aps,twocolumn,prb,preprintnumbers,amsmath,amssymb,superscriptaddress,floats]{revtex4}

\usepackage{graphicx}
\usepackage{bm}
\usepackage{hyperref}
\usepackage{IEEEtrantools}

\usepackage{natbib}
\usepackage{float}

\restylefloat{table}

\def\Cl{$\kappa$-HgCl}
\def\CuCN{$\kappa$-(BEDT-TTF)$_2$Cu$_2$(CN)$_3$}

\begin{document}

\title{Charge and spin interplay in a molecular-dimer-based organic Mott insulator}

\author{Natalia Drichko}\email{Corresponding author: drichko@jhu.edu}
\affiliation{Institute for Quantum Matter and Department of Physics and Astronomy, Johns Hopkins University, Baltimore, MD 21218, USA}
\affiliation{The Institute for Solid State Physics, The University of Tokyo, Kashiwa, Chiba 277-8581, Japan}

\author{Shiori Sugiura}
\affiliation{National Institute for Materials Science, Tsukuba, Ibaraki 305-0003, Japan}
\affiliation{Graduate School of Pure and Applied Sciences, University of Tsukuba, Tsukuba, Ibaraki 305-8577, Japan}

\author{Minoru Yamashita}
\affiliation{The Institute for Solid State Physics, The University of Tokyo, Kashiwa, Chiba 277-8581, Japan}

\author{Akira Ueda}
\affiliation{The Institute for Solid State Physics, The University of Tokyo, Kashiwa, Chiba 277-8581, Japan}
\affiliation{Department of Chemistry, Faculty of Advanced Science and Technology, Kumamoto University, Kumamoto 860-8555, Japan}

\author{Shinya Uji}
\affiliation{National Institute for Materials Science, Tsukuba, Ibaraki 305-0003, Japan}
\affiliation{Graduate School of Pure and Applied Sciences, University of Tsukuba, Tsukuba, Ibaraki 305-8577, Japan}

\author{Nora Hassan}
\affiliation{Institute for Quantum Matter and Department of Physics and Astronomy, Johns Hopkins University, Baltimore, MD 21218, USA}

\author{Yoshiya Sunairi}
\affiliation{The Institute for Solid State Physics, The University of Tokyo, Kashiwa, Chiba 277-8581, Japan}

\author{Hatsumi Mori}
\affiliation{The Institute for Solid State Physics, The University of Tokyo, Kashiwa, Chiba 277-8581, Japan}

\author{Elena I. Zhilyaeva}
\affiliation{Institute of Problems of Chemical Physics RAS, Chernogolovka, Moscow region, 142432 Russia}

\author{Svetlana Torunova}
\affiliation{Institute of Problems of Chemical Physics RAS, Chernogolovka, Moscow region, 142432 Russia}

\author{Rimma N. Lyubovskaya}
\affiliation{Institute of Problems of Chemical Physics RAS, Chernogolovka, Moscow region, 142432 Russia}

\begin{abstract}

Triangular lattice quasi-two-dimensional Mott insulators based on BEDT-TTF molecule and its analogies present a possibility to produce exotic phases by coupling charge and spin degrees of freedom. In this work we discuss magnetic properties of one of such materials, $\kappa$-(BEDT-TTF)$_2$Hg(SCN)$_2$Cl, which is found at the border of the phase transition between a Mott insulator into a charge ordered state. Our magnetic susceptibility and cantilever magnetisation measurements demonstrate how the charge degree of freedom defines magnetic properties for few different charge phases observed in this material as a function of temperature. Between $T_{CO}=30~K$ and $T_S=24~K$ we observe charge and spin separation due to one-dimensional charge stripes formed in this material below $T_{CO}=30~K$. Below $T_S=24~K$ charge and spin degrees of freedom demonstrate coupling. Spin singlet correlations develop below 24~K, however melting of charge order below 15~K prevents the spin singlet state formation, leaving the system in the inhomogeneous state with charge ordered spin singlet domains and charge and spin fluctuating ones.

\end{abstract}

\date{\today}
\maketitle

\section{Introduction}

Research in frustrated magnetism is already for some time focused on a search  for a quantum spin liquid~\cite{Savary2016,Broholm2020}.  The simplest models of this state consider Heisenberg interactions and magnetic or lattice frustration, where triangular lattice is the simplest frustrated lattice. Dimensionality of magnetic system is another parameter defining a behaviour of a frustrated system~\cite{Diep2013frustrated}.  Additionally, there are theoretical predictions for obtaining this quantum states by coupling spins to other fluctuating degrees of freedom. Example of that are spin-orbital liquids, where spins are coupled to fluctuating orbital degrees of freedom~\cite{Balents2010}, or quantum dipole liquids~\cite{Hotta2010,Naka2016,Yao2018}. The latter state can be realized in organic Mott insulators, where on-site fluctuating electrical dipole moments are coupled to the spin degree of freedom ~\cite{Hotta2010,Naka2016, Yao2018}. Some experimental results suggest a realization of these fluctuating dipole moments, and associated with them   quantum dipole liquid state\cite{Hassan2018,Hassan2020}. An experimental evidence of coupling of electric dipoles to the spin degree of freedom is necessary to prove that this path to a spin liquid state is possible at all.

Studies of the $\kappa$-(BEDT-TTF)$_2$Hg(SCN)$_2$Cl (\Cl) in the insulating state can provide the answer to this question.   This material is known to form a ferroelectric order of electric dipoles at temperatures between $T_{CO}=30~K$ and 15~K \cite{Drichko2014,Gati2018,Hassan2020}, which melts on cooling to the lower temperatures. NMR measurements suggest the  absence of spin order in this material down to 25~mK~\cite{Pustogow2020}. Using SQUID magnetic susceptibility, and cantilever torque magnetometery measurements we follow  the temperature evolution of magnetic properties in \Cl\ from metallic into dipole ordered state, and further into the melting of the dipole order. We observe the evidence of decoupled charge and spin degrees of freedom below $T_{CO}=30~K$ the temperature of the transition into the charge ordered state characterized by charge stripes, while below $T_S=24~K$  the coupling of charge and spin degrees of freedom defines magnetic properties.  No magnetic order is detected in this system down to at least 120 mK, while the exchange interactions are on the order of at least 100~K.

\begin{figure}
	\includegraphics[width=9cm]{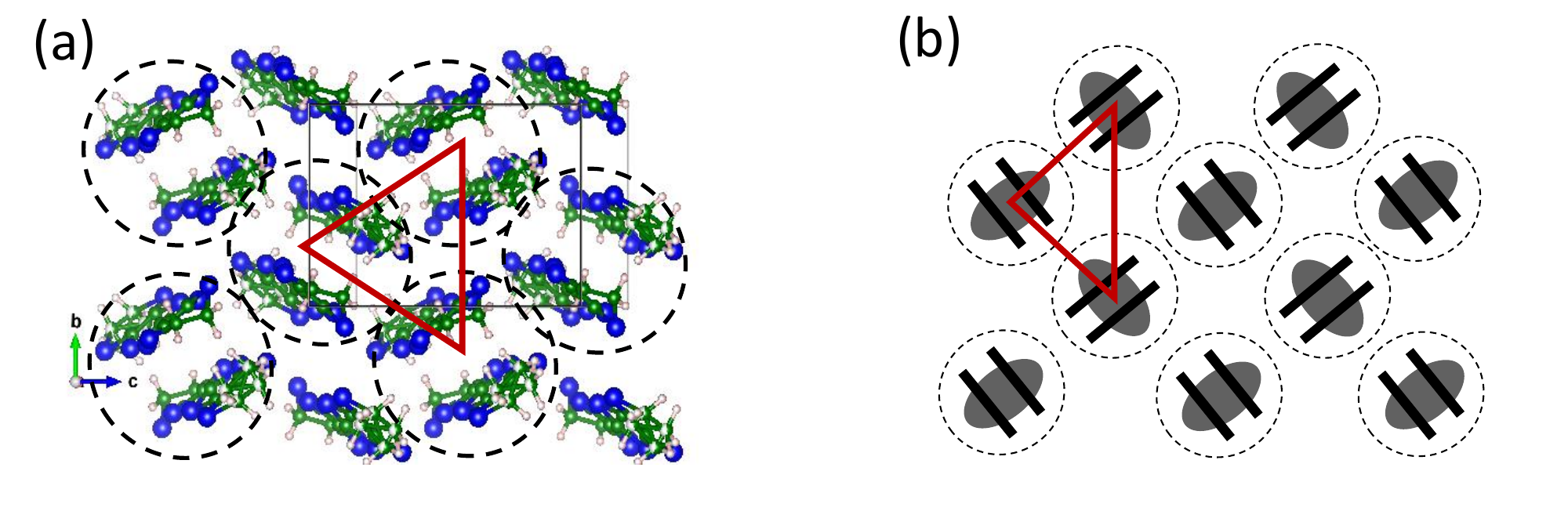}
	\caption{(a) A projection of the structure of (BEDT-TTF)$_2$ layer of   $\kappa$-(BEDT-TTF)$_2$Hg(SCN)$_2$Cl on $bc$ crystallographic plane. Dashed line circles indicate dimers of (BEDT-TTF)$_2$.  (b) A schematic representation of the 2D layer of (BEDT-TTF)$_2$ dimers. Thick black lines represent BEDT-TTF molecules bound in a dimer by a shared orbital (grey oval). In both (a) and (b) the red triangle show a unit of triangular lattice. }
	\label{fig_0}
\end{figure}

The crystal structure of \Cl, which allows electrical dipole formation on the lattice sites, with S=1/2 associated with each site is shown in Figs.~\ref{fig_0} and \ref{fig_torque}a.  In the layered crystal structure of this material,  (BEDT-TTF$)_2^{1+}$ dimers form a triangular lattice in the BEDT-TTF-based layer, with  charge $+1e$ and S=1/2 per dimer  achieved by a charge transfer between anion and cation layers.
On the charge order transition, a  charge difference of $\Delta n$=0.2$e$ between charge-poor and charge rich molecules of a dimer leads to a dipole solid state~\cite{Drichko2014,Gati2018}. This charge distribution is schematically indicated by the red and blue colors in the  (BEDT-TTF)$_2$ dimers in Fig.~\ref{fig1_vX}.  The charge order transition drives  a large increase of d.c. resistivity~\cite{Yasin2012,Drichko2014} and a notable feature in heat capacity (Fig.~\ref{fig1_vX})~\cite{Hassan2020,Gati2018}. While the lattice response is detected on this first order phase transition~\cite{Gati2018},  no structural phase transition   has   been observed by the XRD studies so far~\cite{Drichko2014}.

On the charge ordering transition the electronic system changes dimensionality from 2D to  1D due to the formation of the charge stripes along the $c$ axis~\cite{Drichko2014,Hassan2020,Gati2018}. Calculations confirm that the charge order in \Cl\  results in a high in-plane anisotropy of magnetic exchange interactions $J$, leading to a suggestion of an effectively 1D magnetism for \Cl~\cite{Jacko2020}. Two scenarios for the ground state of these effectively 1D antiferrmagetic (AF) chains are possible: In the absence of coupling to the lattice, the 1D  stripes can evolve into a 1D spin liquid, presenting another possible way to reach a spin liquid state for \Cl~\cite{Mila2000,Watanabe2017}. A strong coupling of spin degrees of freedom to the lattice can result in a  formation of a spin singlet state~\cite{Dayal2011}, as observed, for example, in $\theta$-(BEDT-TTF)$_2$RbZn(SCN)$_4$~\cite{Mori1997,Kanoda2011}.

Melting of the charge order in \Cl\ below 15~K~\cite{Hassan2020} can lead to the charge disorder and an appearance of  domain walls. An interplay of disorder and low-dimensionality is an important direction of research for spin liquid materials. Some materials, a recent example of which is  a  triangular lattice magnet  YbMgGaO$_4$, have the intrinsic structural disorder control of the magnetic ground state~\cite{Kimchi2017}.  In order to study structural disorder effects in a controlled fashion, disorder is introduced by X-ray irradiation~\cite{Furukawa2015} or doping~\cite{Kawamata2008}. \Cl\ provides  a unique situation where charge inhomogeneities are present only in the  temperature range below 15~K, and their scale is temperature dependent.

Following magnetic properties of \Cl\ below 50~K in these different charge phases allows us to demonstrate how the charge degree of freedom defines magnetic properties of the system. Charge order below $T_{CO}=30~K$ resulting in 1D charge stripes leads to a decoupling of charge and spin degrees of freedom in the limited temperature range. Spin singlet correlations in the chains develop at $T_S=24~K$, however the long range spin singlet formation is prevented by the charge order melting below 15~K. As a result, \Cl\ shows no long range magnetic order down to 120~mK, and can be understood as an   inhomogeneous a mix of spin-singlet and spin fluctuating domains, where the inhomogeneous state is a consequence of the system being close to the phase border with the ferroelectric state~\cite{Naka2010}.

\section{Experimental}

Single crystals of $\kappa$-(BEDT-TTF)$_2$Hg(SCN)$_2$Cl (\Cl)  were prepared by electrochemical oxidation of the BEDT-TTF solution in 1,1,2-trichloroethane (TCE) at a temperature of 40$^{\circ}$ C and a constant current of 0.5~$\mu$A. A solution of Hg(SCN)$_2$, [Me$_4$N]SCN·KCl, and dibenzo-18-crown-6 in 1:0.7:1 molar ratio in ethanol/TCE was used as supporting electrolyte for the \Cl\ preparation.  The composition of the crystals was verified by electron probe microanalysis and X-ray diffraction.

Temperature dependence of heat capacity in the temperature range from 50~K to 200~mK was measured using Quantum Design PPMS system equipped with the DR the option for crystals of the mass 2-4 mg. Temperature dependence of magnetic susceptibility $\chi_M(T)$ of a polycrystal sample of m=2.451~mg was measured using Quantum Design MPMS in field of 1~T in the temperature range from 2 to 300~K. Magnetic torque of single crystals was measured using magnetic cantilever setup in the temperature range down to 120~mK and magnetic fields up to 18~T. The setup and the details of the data analysis are discussed in \cite{Yamashita2010a}.

\section{Results}

\begin{figure}[h]
	\includegraphics[width=8cm]{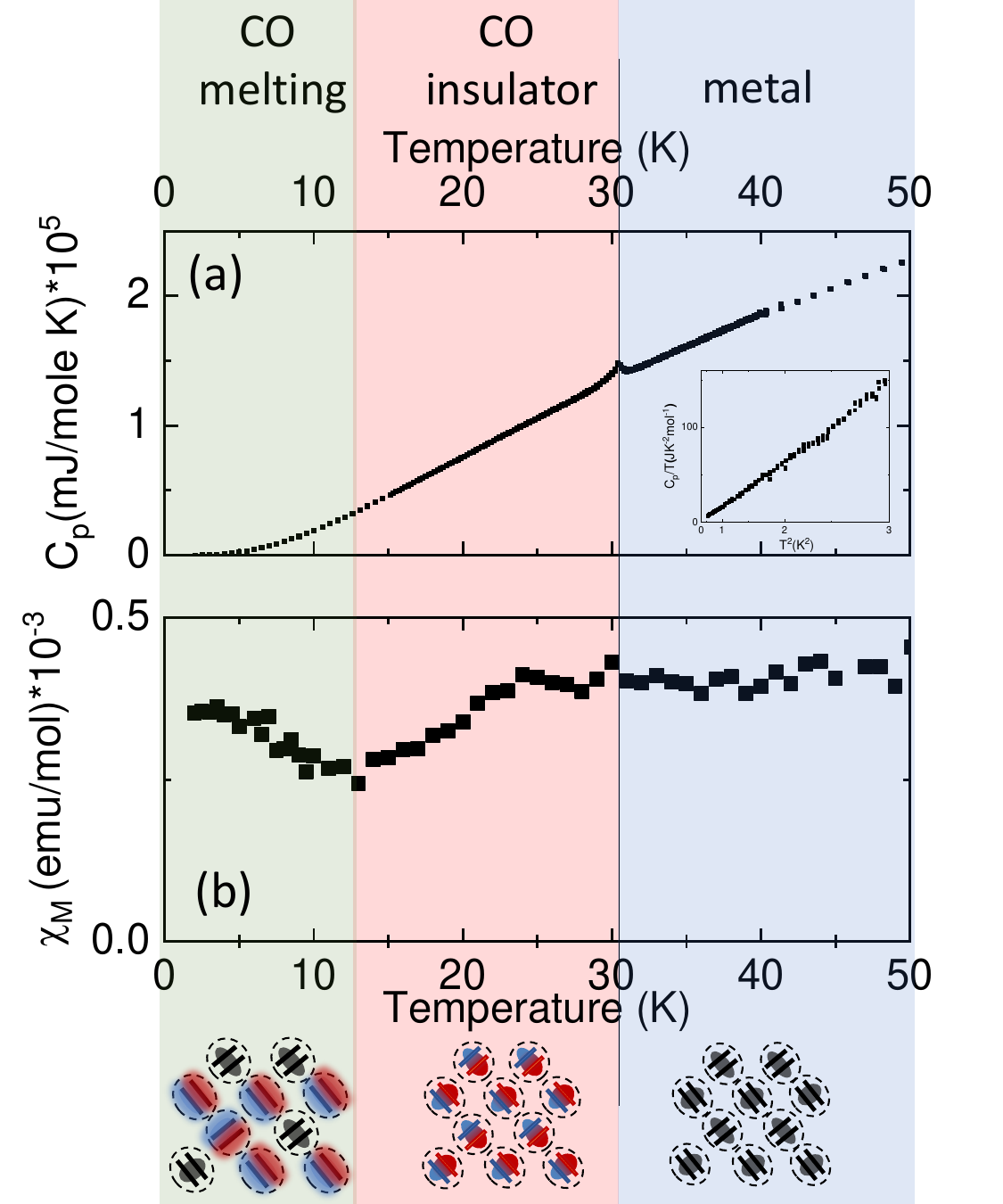}
	\caption{ (a) Temperature dependence of the specific heat C$_p$(T) of $\kappa$-(BEDT-TTF)$_2$Hg(SCN)$_2$Cl at 0~T. The inset shows heat capacity in the range 200~mK - 3~K. Note $\gamma$= 0 within the error of the measurements. (b) Temperature dependence of  $\chi_M (T)$.  Three regimes related to the charge degree of freedom~\cite{Hassan2020} are indicated by color. Schematic charge distribution on the orbital of (BEDT-TTF)$_2$ dimer is shown, where grey indicates homogeneous charge in the metallic state above $T_{CO}=30~K$, red is charge-rich, and blue is charge-poor molecules in the charge ordered insulating state. Low temperature mixed state below 15~K consists of components with homogeneous charge distribution, and dimers where charge order is preserved as static or slowly fluctuating, as depicted by blurred colors.}
	\label{fig1_vX}
\end{figure}

On the metal-insulator transition in \Cl\ at $T_{CO}=30~K$, heat capacity shows a distinct peak (see Fig.~\ref{fig1_vX} (a)). No other phase transition is detected down to 200~mK (see inset in Fig~\ref{fig1_vX}). The extrapolation down to 0~K suggests a  negligible linear component of $\gamma$ term in heat capacity $C_p=\beta T^3 + \gamma T$ (see Supplemental Material (SM)).

Magnetic susceptibility  $\chi_M (T)$ of a polycrystal sample of  2.451~mg  measured at 1 T (Fig.~\ref{fig1_vX} (b)) reveals a complex temperature dependence. Above $T_{CO}=30~K$ \Cl\ shows Pauli susceptibility of about 4$\cdot$10$^{-4}$~emu/mol. This value is close to that of the other BEDT-TTF-based organic conductors~\cite{Sekretarczyk1988}.
At the charge order transition magnetic susceptibility $\chi_M (T)$ does not show any change  within the noise of the measurements. Instead,  magnetic susceptibility starts to decrease abruptly on cooling below about 24~K.~\footnote{A seemingly similar material $\kappa$-ET$_2$Cu$_2$(CN)$_3$ that shows a very close magnetic susceptibility values of  $\chi_M$= 5*10$^{-4}$~emu*mol$^{-1}$ at 30~K~\cite{Shimizu2003} is not the right compound to compare to. This system shows magnetic excitations spectrum of a two-dimensional triangular lattice of ET-dimers, and lacks any charge disproportionation. However, in any case, $\chi_M (T)$ of \Cl\ decreases faster than a triangular lattice S=1/2 magnetic susceptibility.~\cite{Sedlmeier2012}}.
Decreasing susceptibility in the temperature range between 24~K and 15~K can be fit by  a magnetic susceptibility of a gapped system ~$e^{-\frac{\Delta}{k_BT}}$, which yields a gap $\Delta$ =36~K=1.5~T$_c$ (see Appendix) with the assumption that susceptibility of a charge ordered system will become $\chi_{CO}=0$ on cooling.  Magnetic susceptibility starts to rise again on cooling the sample below 15~K, and saturates below about 5 K, with the saturation values of about  3.5$\cdot$10$^{-4}$ emu/mol.  No indication of magnetic ordering  is found in $\chi_M (T)$ of \Cl\   down to 2~K.

BEDT-TTF-based crystals are typically very small, and posses very low magnetic susceptibility (Fig~\ref{fig1_vX}). In order to understand the nature of the low temperature magnetic state, we performed measurements of the cantilever torque magnetization for single crystals of \Cl. This method proved to be the most sensitive to detect magnetic order, and was successfully applied to organic Mott insulators~\cite{Pinteric1999,Isono2016,Watanabe2012}.   Magnetic torque signal measured for \Cl\  is described well by the following equation:

 \begin{equation}
 \tau = \tau_0 + \tau_{\theta} \mathrm{sin} (\theta - \theta_1) + \tau_{2\theta} \mathrm{sin} 2(\theta - \theta_2)
 \end{equation}

In \Cl\ torque response,  $\tau_{\theta} \mathrm{sin} (\theta - \theta_1)$ component does not change with the applied magnetic field at all measured temperatures (see Appendix). We conclude that it fully corresponds to the gravity force, no ferromagnetic component of  torque was detected. The $\tau_{2\theta} \mathrm{sin}(2(\theta - \theta_2))$ component in the torque response of \Cl\ corresponds to the paramagnetic response.

\begin{figure}[h]
	\includegraphics[width=9cm]{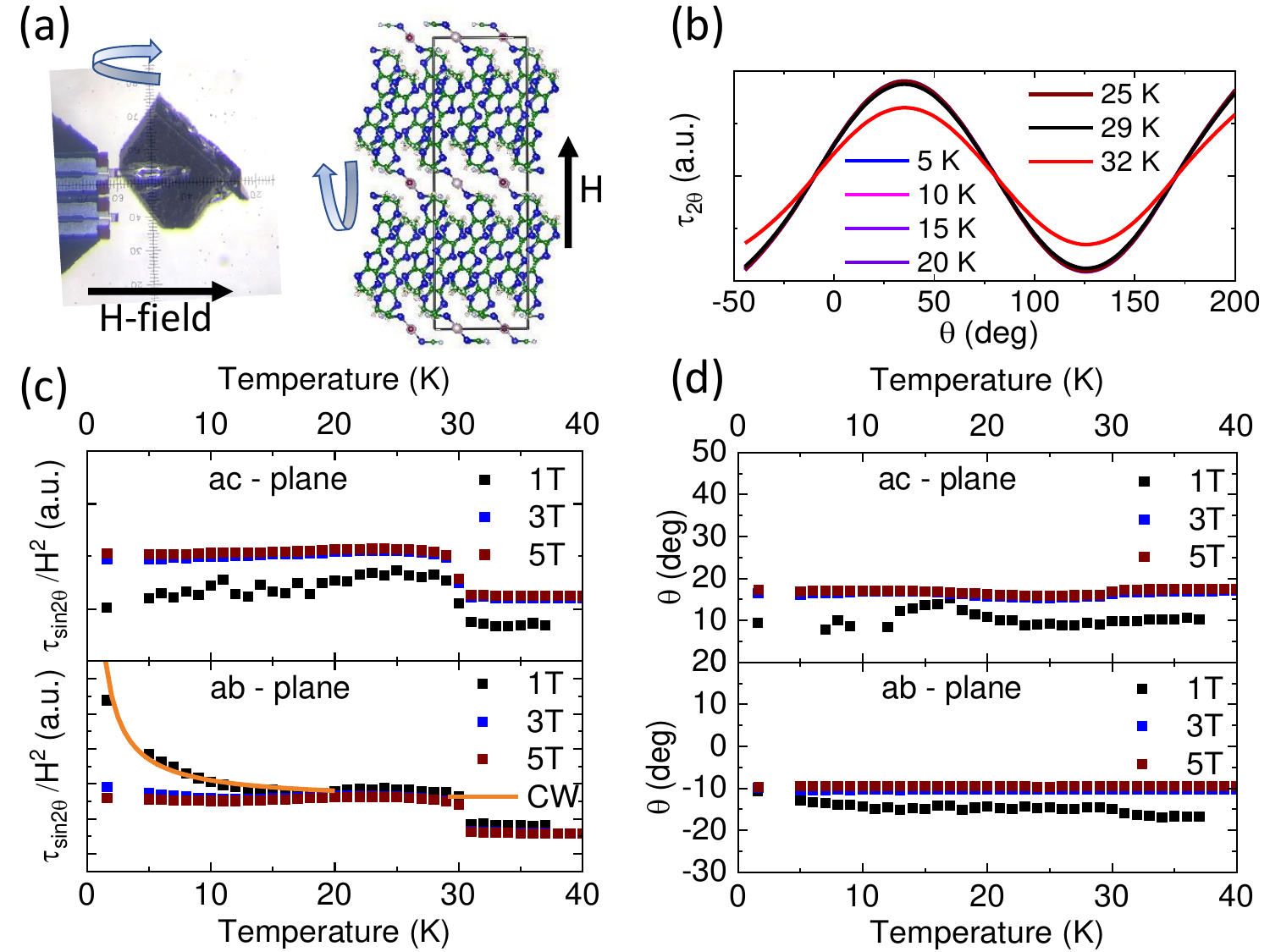}
	\caption{(a) Scheme of cantilever torque magnetometery experiment with the rotation direction for the $ab$ plane shown with blue arrow, and magnetic field direction shown in black. (b) Angle dependence of paramagnetic component of torque $\tau_{2\theta}$ at 5~T, rotation in the $ab$ plane for temperatures related to the different charge states of $\kappa$-(BEDT-TTF)$_2$Hg(SCN)$_2$Cl. (c) Temperature dependence  of the paramagnetic torque amplitude $\tau_{2\theta}$ in $ac$ plane and $ab$ planes. (d) Temperature dependence of the phase $\theta_2$ in the  $ac$ and $ab$ planes. }
	\label{fig_torque}
\end{figure}

Cantilever torque magnetization measurements show the persistence of  the paramagnetic  response when \Cl\ is cooled through the charge order transition at 30~K and below this temperature, but detect an abrupt increase of torque amplitude  $\tau_{2\theta}$ at $T_{CO}=30~K$, as shown in Fig.~\ref{fig_torque}. Torque amplitude  $\tau_{2\theta}$ for the rotation in the $ab$ plane  at 1 T roughly follows the temperature dependence of magnetic susceptibility, with a decrease at about 20~K, and an increase below 15~K. The phase $\theta_2$ for the rotation in the $ac$ plane follows this temperature behavior as well, while the amplitude stays constant on cooling.  These effects are weak,  and are suppressed at 3~T.

\begin{figure}
	\includegraphics[width=9cm]{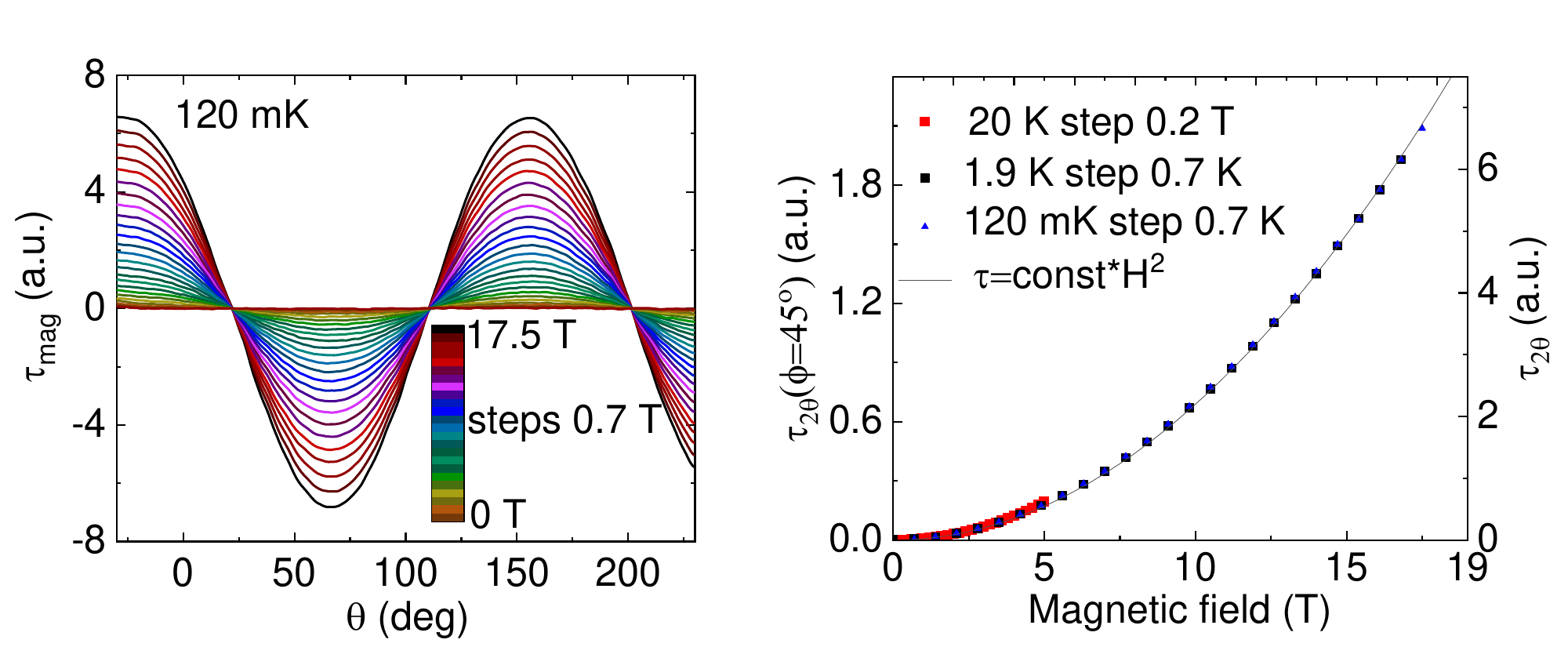}
	\caption{Magnetization of  $\kappa$-(BEDT-TTF)$_2$Hg(SCN)$_2$Cl measured my magnetic cantilever torque magnetometery. (a) Magnetic field dependence of torque signal at 120~mK for magnetic field $H$ in the range from 0 to 17.5~T.  (b) Field dependence of $\tau_{2\theta}$.}
	\label{fig:LowT}
\end{figure}

Magnetization is probed by measuring the  torque amplitude dependence on magnetic field $H$ at an angle where the amplitude of the torque is maximum, see Fig.~\ref{fig:LowT}. At temperatures T=20, 10, 5, 1.9~K for the field up to $H = 5~T$ torque $\tau$  shows parabolic dependence  $\tau \propto H^2$, suggesting a purely paramagnetic state.
The $\tau \propto H^2$ behavior of magnetization was confirmed at T=120 mK in magnetic field up to 17.5~T (Fig.~\ref{fig:LowT}).

\section{Discussion}

A picture of an exotic spin and charge behavior in  \Cl\ forms when we put together the results obtained with these different probes. An absence of a change of magnetic susceptibility at  $T_{CO}=30~K$ and purely paramagnetic torque response below the transition are unexpected: The character of $\chi_M (T)$ should change from Pauli to a behavior of an  insulator with unpaired spins or a spin singlet state.  However, the charge ordering transition only leads to a jump in the paramagnetic component of the torque amplitude $\tau_{2\theta}$ at 30~K, which is a measure  the anisotropy of  g-factor~\cite{Uji1997} and responds to the change of the electronic structure on the transition.

$\chi_M (T)$ starts to  decrease on cooling only below $T_S=24~K$, suggesting a development of spin singlet correlations. This decrease it  in agreements with the previous results of ESR measurements~\cite{Yasin2012}. Heat capacity  does not show any indication of a phase transition at 24~K, but  it is known to be not a sensitive indicator of a magnetic transition neither in 1D~\cite{deSousa2009}, nor in 2D~\cite{YamashitaS2010} organic-based systems.


The development of a spin singlet state is one of the expected consequences of the charge stripes formation~\cite{Dayal2011}. Striking here is a decoupling of the charge order transition temperature (T$_{CO}=30~K$) and the temperature of the spin singlets formation $T_S=24~K$. It demonstrates the decoupling of charge and spin degrees of freedom in this temperature range as a consequence of 1D electronic structure~\cite{Drichko2014,Gati2018,Hassan2020} of charge stripes in \Cl. Such a decoupling is common in  1D TMTTF -based materials, which undergo a transition into an antiferromagnetically ordered or spin-Peierls state at temperatures of about ten times lower than the temperature of a metal-insulator transition~\cite{Salameh2011,Dressel2007}. To the best of our knowledge, it is observed for the first time in a layered system as a result of a dimensional crossover associated with 1D charge stripes formation.

A signature of a singlet state in magnetic torque would be a decrease of torque amplitude $\tau_{2\theta}$ down to zero. We observe that  magnetic torque of \Cl\ preserves its paramagnetic character below 24~K and shows only a  slight decrease of the torque amplitude $\tau_{2\theta}$  in the $ab$ plane. A decrease of $\chi_M$ without a detectable long range order suggests that AF order or spin singlet pairs  which will posses $\chi_{CO}$=0 are formed with a certain correlation length in the temperature range between 24 and 15~K, without long range order. Changes of the phase values of torque in the $ac$ plane observed at 1~T  (Fig. 2)  follow the susceptibility behavior and are apparently due to orphan spins appearing as a result of short range AF or spin single correlations~\cite{Riedl2019}.

At temperatures below 15~K, the charge order starts to melt~\cite{Hassan2020}, and  magnetic susceptibility increases again. In the resulting inhomogeneous system depicted schematically in Fig.~\ref{fig1_vX} the charge order melted fraction $\rho$  would provide $\chi_M$= $\rho \chi_{H}$ +  $\chi_D$, where  $\chi_H$ corresponds to the response of the charge melted fraction with finite susceptibility, and $\chi_D$ is the susceptibility of domain walls between  fractions with $\chi_H$ and $\chi_{CO}$=0. According to our previous results, the charge melted fraction  $\rho$  increased on cooling from $\rho = 0$ in the charge ordered state to $\rho = 1/3$ at 2~K~\cite{Hassan2020}. The increase of both $\rho \chi_{H}$ and  $\chi_D$ components on melting of the charge order  can add to the increase  of the total $\chi_M$  of the sample as the temperature is lowered from 15 to 2~K.

The system where charge order is melting is intrinsically inhomogeneous. If paramagnetic spins of inhomogeneities are non-interacting, they would  produce Curie-like response~\cite{Ami1995}. However, the increase of magnetic susceptibility below 15~K differs from a simple Curie-like response: Instead of diverging as 1/T, $\chi_M (T)$ shows a weaker temperature dependence and flattens below 5~K. The increase of torque amplitude $\tau_{2\theta}$ is observed only in the $ab$ plane, showing that the system preserved some anisotropy, and is also weaker than pure Curie response. We can conclude, that  paramagnetic spins of the melted charge order and domain walls are interacting and present a more complicated picture than orphan spins of defects (Ref.~\onlinecite{Riedl2019} and references therein). Recently is was suggested that two spins on the ends of fluctuating charge order chains interacting through a non-charge-ordered dimer could interact ferromagnetically~\cite{Yamashita2020}. It worth noting, that the increase and saturation of $\chi_M(T)$  below 15~K in \Cl\  is  similar to the dipole liquid candidate $\kappa$-(BEDT-TTF)$_2$Hg(SCN)$_2$Br magnetic susceptibility, but on a much smaller scale.   Other models developed for quantum paraelectrics also suggest an increase of magnetic susceptibility due to fluctuating electric dipoles~\cite{Dunnett2019}.


If the increase of the $\chi_M (T)$ on cooling and  the temperature dependence of torque at low fields are produced by the interacting domain walls, they become fully polarized already at 3 ~T according to the torque measurements. This result is in agreement with a suppression of a peak in T$^{-1}$ in NMR~\cite{Pustogow2020} response by the magnetic field directed perpendicular to BEDT-TTF-based layers, and suggests that the origin of the peak is the response of the domain walls. According to the Raman data in magnetic field in this direction, the charge distribution itself does not change in field up to 30~T~\cite{Hassan2021}.

No indication of magnetic ordering  is observed in the magnetic cantilever torque studies of \Cl\ down to 120 mK. The paramagnetic torque amplitude $\tau_{2\theta} \propto H^2$ up to   17.5~T, and the phase is constant with field.
This shows the absence of magnetic order or of a spin singlet state in the fraction of the  \Cl\ crystal where charge order is melted and lattice cites possess  one unpaired electron (S=1/2) with estimated  magnetic exchange $J$ of the order of 100~K or higher~\cite{Yamashita2020,Jacko2020,Hassan2018}. This part of the system provides the response suggestive of a spin liquid state from NMR measurements~\cite{Pustogow2020}.

The low temperature magnetic state of both charge ordered and charge melted fractions in \Cl\ has signatures very different from the other spin liquid candidates triangular 2D dimer Mott insulators~\cite{Shimizu2003,Yamashita2008,Yamashita2011,Yamashita2010,Pratt2011,Isono2014}.
 In S=1/2 triangular lattice organic  Mott insulators without a  charge degree of freedom, such as \CuCN~\cite{Nakamura2014,Hassan2018}, $\kappa$-(BEDT-TTF)$_2$Ag$_2$(CN)$_3$~\cite{Nakamura2017}, and Pd(dmit)$_2$-based materials~\cite{Nakamura2015}, Raman scattering spectroscopy detects  a continuum of magnetic excitations well understood in terms of S=1/2 on a triangular lattice~\cite{Perkins08,Vernay2007,Holt2014,Hassan2018Crystals}.   This continuum is absent in the Raman scattering spectra of  \Cl~\cite{Hassan2018} down to 10~K, and  magnetic excitations in this material are still to be detected.
In addition, in contrast to dimer Mott insulators without charge order~\cite{Yamashita2011, Yamashita2010a}, \Cl\ shows   $\gamma$=0 linear term in the heat capacity~\cite{Hassan2018}, suggesting that magnetic excitations spectrum is gapped, which can be a consequence of a finite size of domains with fluctuating spins. Apparently, the  inhomogeneities and domain walls do not produce the low temperature input in $\gamma$.

The key difference between the electronic structure of \Cl\ and the triangular 2D  Mott insulators discussed above is the smaller intradimer integral in \Cl, which results in a weaker dimerization and active charge degree of freedom~\cite{Gati2018,Hassan2020,Naka2010}. The relevant phase diagram of the $\kappa$-phase organics with the intra dimer transfer integral serves as a tuning parameter, tuning a system from an antiferromagnetic Mott insulator at high dimerization to a charge order ferroelectric at low dimerization is theoretically discussed in Ref.~\cite{Naka2010}. We have already shown that \Cl\ is found on the border between these two phases and experiences a re-entrant charge order melting transition due to the   competition of electronic repulsion which leads to charge order and antiferrmagnetic interactions ~\cite{Hassan2020,Naka2010}. The formation of domains at low temperature in this material is a consequence of the competition of these parameters close to phase border. In this work we demonstrate, that the ordered vs fluctuating domains are found not only in the charge sector, but also in the spin degree of freedom.

\section{Conclusions}

In conclusion, our data experimentally  demonstrate the importance of the  interplay of the charge and spin degrees of freedom for the dimer Mott insulators, previously suggested theoretically~\cite{Hotta2010,Naka2010}. We reveal  how the charge state in \Cl\ in  each temperature range  defines the magnetic properties.
On cooling we first  observe a spin-charge separation  in the narrow temperature range between $T_{CO}=30~K$ and $T_S=24~K$ due to the formation of 1D charge stripes. The effects at lower temperatures demonstrate coupling between charge and spin degrees of freedom. Below $T_S=24~K$, the system shows spin singlet correlations, but preserves considerable paramagnetic torque response, proving that these correlations never  develop into a long range order. Moreover, below 15~K the melting of the charge order prevents the spin singlet formation in the whole system, and leads to an inhomogeneous state with spin singlet charge ordered domains, and domains which do not show either charge or magnetic order.


\section{Acknowledgements}

The authors are grateful to C. Broholm and T. Clay for fruitful discussions. Work at JHU was supported by NSF award DMR-2004074.
The work in Chernogolovka was performed in accordance with the state task,
state registration No. AAAA-A19-119092390079-8. The work  in  Japan  was  supported  by  KAKENHI  (Grants-in-Aid   for   Scientific   Research)   Grants   No.JP17K05533,No.    JP18H01173,  No.    JP17K05497,  No.    JP17H02916,JP19K05397, 19H01848, and 19K21842. ND is grateful for the travel support from the IPMU-ISSP-JHU collaborative program in Physics and Astronomy.

\bibliography{OrganicConductors2020}

\newpage

\section{Appendix}

{\bf Magnetic susceptibility}

\begin{figure}
	\includegraphics[width=7cm]{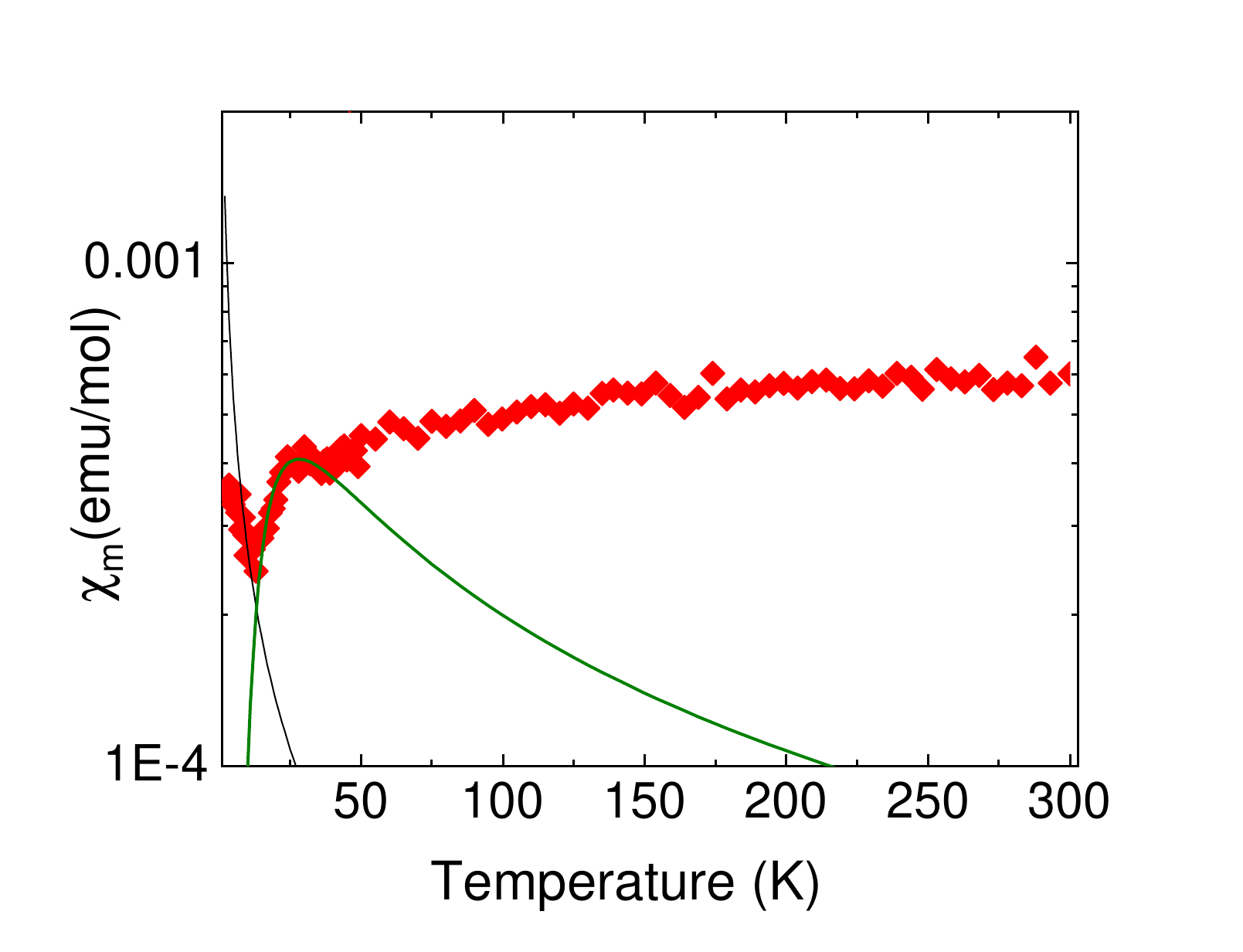}
	\caption{Magnetic susceptibility  of  $\kappa$-(BEDT-TTF)$_2$Hg(SCN)$_2$Cl polycrystal sample  measured by SQUID magnetometer using H=1~T in the temperature range from 300 to 2 K.}
	\label{SI_chi}
\end{figure}

Fig.~\ref{SI_chi} shows magnetic susceptibility data in the whole measured range. The figure also demonstrated a curve of $\chi_{dimer} = Ae^{-\frac{\Delta}{k_BT}}$ which can describe $\chi_M(T)$ in the temperature range between 24 and 15~K, and $\chi_{Curie}$=C/T to compare to the low-temperature behavior of $\chi_M(T)$.

{\bf Cantilever magnetic torque measurements}

\begin{figure}[h!]
	\includegraphics[width=7cm]{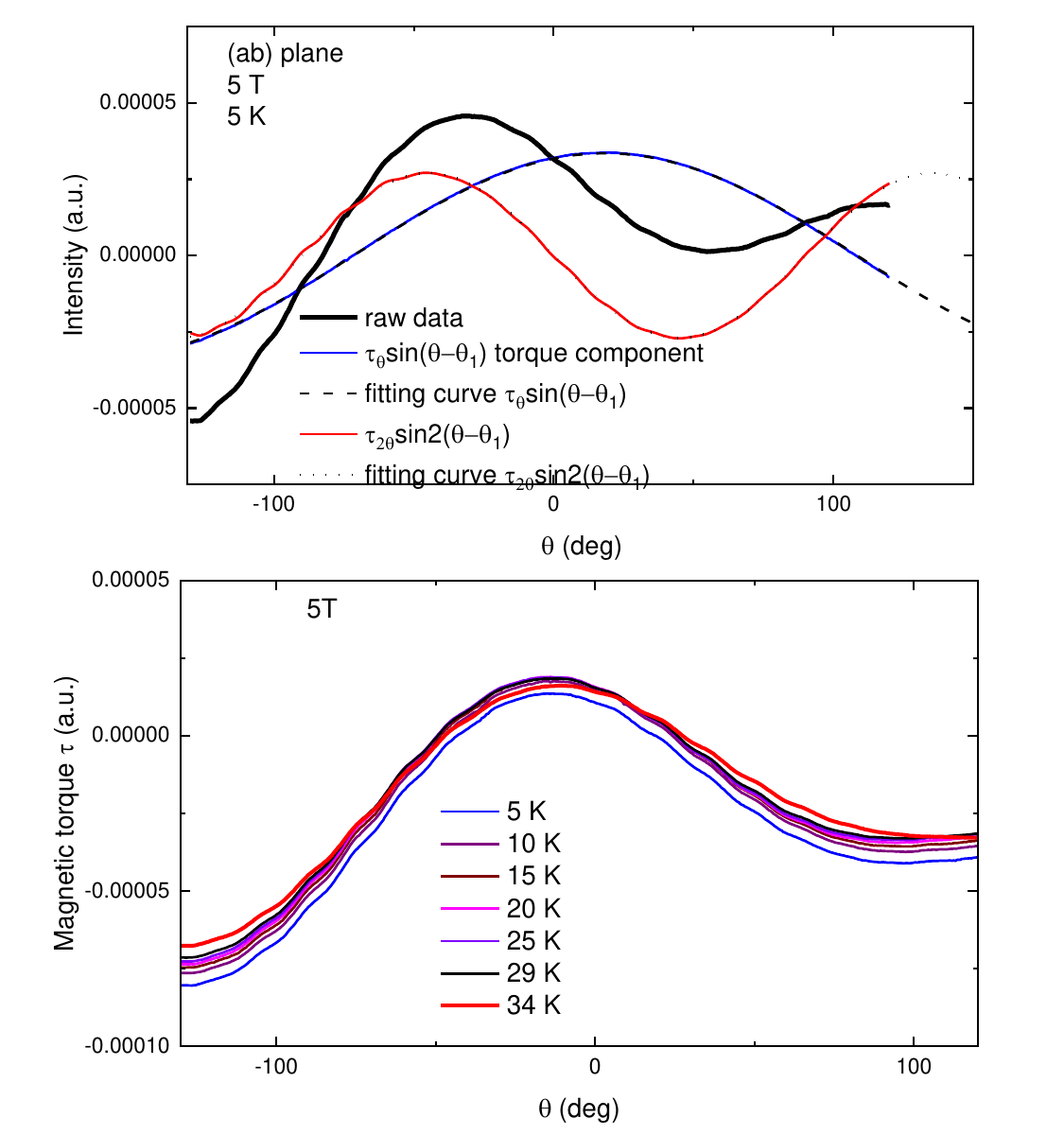}
	\caption{Upper panel: Cantilever magnetic torque data analysis, and example for the data for the rotation in $ab$ plane at 5~K in magnetic field of 5~T. The figure shows original data, and $\tau_{\theta}$sin($\theta-\theta_1$), and  $\tau_{2\theta}$sin2($\theta-\theta_1$) components with the respective fitting curves. Lower panel: Raw data  of $\kappa$-(BEDT-TTF)$_2$Hg(SCN)$_2$Cl obtained by cantilever magnetic torque measurements. Measurements are done for rotation in $ab$ plane, H= 5T. Angle dependence of $\tau_{2\theta}$ extracted from these measurements is presented in Fig.~ 3(b)}
\label{SItorque1}
\end{figure}

Fig~\ref{SItorque1} presents an example of analysis of the cantilever magnetic torque data. The full torque $\tau$ (black curve) is reproduced well by the sum of $\tau_{\theta}$sin($\theta-\theta_1$), and  $\tau_{2\theta}$sin2($\theta-\theta_1$) components.  Calibration of the cantilever response by gravity signal is discussed in details in Ref.~\cite{Yamashita2010a}

\end{document}